\documentclass[11pt]{article}
\usepackage{jheppub}
\usepackage{bm} 
\def\n{{\bf n}}

\def\p{{\bf p}}
\def\x{{\bf x}}

\def\grad{{\bm \nabla}}
\def\coeff#1#2{{\textstyle {\frac {#1}{#2}}}}

\def\M{{\cal M}}
\def\L{\pounds}
\def\S{{\rm s}}
\def\aE{\chi_{\rm\scriptscriptstyle E}}

\def\aEE{\chi_{\rm\scriptscriptstyle EE}}
\def\aBB{\chi_{\rm\scriptscriptstyle BB}}
\def\aEB{\chi_{\rm\scriptscriptstyle EB}}
\def\aBE{\chi_{\rm\scriptscriptstyle BE}}

\def\EM{{\rm\scriptscriptstyle EM}}
\def\MB{M_{\scriptscriptstyle B}}
\def\MO{M_{\scriptscriptstyle\Omega}}
\def\gE{g^{\scriptscriptstyle E}}
\def\AE{A^{\scriptscriptstyle E}}

\title{Thermodynamics of polarized relativistic matter}
\author{Pavel Kovtun}
\affiliation{
Department of Physics and Astronomy,
University of Victoria,\\
PO Box 1700 STN CSC,
Victoria BC,  V8W 2Y2,
Canada
}
\abstract{
We give the free energy of equilibrium relativistic matter subject to external gravitational and electromagnetic fields, to one-derivative order in the gradients of the external fields. The free energy allows for a straightforward derivation of bound currents and bound momenta in equilibrium. At leading order, the energy-momentum tensor admits a simple expression in terms of the polarization tensor. Beyond the leading order, electric and magnetic polarization vectors are intrinsically ambiguous. The physical effects of polarization, such as the correlation between the magneto-vortically induced surface charge and the electro-vortically induced surface current, are not ambiguous. 
}

\begin{document}
\maketitle

\section{Introduction}

We would like to understand collective macroscopic behaviour of matter subject to external fields. In the high-temperature limit this is often captured by classical hydrodynamics. The ingredients that go into writing down the hydrodynamic equations are: the identification of relevant variables (conserved densities, order parameters), the derivative expansion (small gradients near equilibrium), and symmetry constraints.
The hydrodynamic equations are modified when the system is subject to external electric and magnetic fields. The latter will induce polarization (electric, magnetic, or both) in a fluid, and as a result the transport properties of the fluid will change. Our focus here will be on isotropic relativistic matter because {\it a)} electromagnetic fields are intrinsically relativistic,  {\it b)} relativistic fluids have more symmetry than non-relativistic fluids, and {\it c)} relativistic fluids have been a subject of much recent attention in the literature due to their appearance in heavy-ion physics~\cite{Jeon:2015dfa, Schafer:2009dj}, in gravitational physics, through the holographic duality~\cite{Rangamani:2009xk}, and even in condensed matter physics~\cite{PhysRevLett.103.025301, Lucas:2016omy}. The systematic description (including the derivative expansion) of polarized relativistic fluids is largely missing, and the present paper is a step in filling that gap. 

In order to understand the hydrodynamics of matter subject to external fields, one needs to understand its thermodynamics first. In what follows we will describe the procedure for obtaining the energy-momentum tensor and the current density for stationary equilibrium polarized matter subject to external gravitational and electromagnetic fields. 
We will find simple expressions for ``bound'' currents, including equilibrium surface currents and surface momenta. 

Let us start with the standard description of equilibrium thermodynamics without external fields. In the grand canonical ensemble at temperature $T_0=1/\beta_0$ and chemical potential $\mu_0$, extensivity in the large-volume limit dictates that the logarithm of the grand-canonical partition function $Z[T_0,\mu_0]$ is proportional to the $d$-dimensional spatial volume, 
$$
  \ln Z = \beta_0 \int\! d^dx\; P(T_0, \mu_0)\,,
$$
where the pressure $P(T_0, \mu_0)$ is constant in equilibrium~\cite{LL5}. The partition function $Z$ may be computed from a Euclidean path integral with a Euclidean (imaginary) time compactified with period $\beta_0$, see e.g.~\cite{Brown-QFT}. In the path integral action, the fundamental fields of the microscopic theory can then be coupled to time-independent external sources: the (Euclidean) metric $\gE_{\mu\nu}$ and the (Euclidean) gauge field $\AE_\mu$. The gauge field couples to the conserved current, whose time component is the charge density corresponding to the chemical potential. See Ref.~\cite{Banerjee:2012iz} for a convenient parametrization of the Euclidean sources $\gE$ and $\AE$. The Euclidean path integral  gives rise to the partition function $Z=Z[T_0,\mu_0, \gE, \AE]$, where $T_0=1/\beta_0$ is the coordinate periodicity of the Euclidean time. We assume that the coupling to time-independent external sources leaves the system in equilibrium, so that no entropy is produced. The temperature and the chemical potential will be altered by the external sources and are not uniform any more. For example, the equilibrium temperature becomes $T(\x)=T_0/\sqrt{\gE_{00}(\x)}$~\cite{LL5}. Similarly, the chemical potential will be shifted by the time component of the external gauge field. We can write $W=-i\ln Z$ as
\begin{equation}
\label{eq:WE}
  W[T_0,\mu_0, \gE, \AE] = -i\beta_0 \int\!d^dx\; 
  \sqrt{\gE}\, {\cal F}(T_0,\mu_0, \gE, \AE)\,,
\end{equation}
where $\sqrt{\gE}$ is the square root of the determinant of $\gE_{\mu\nu}$, and ${\cal F}$ is the negative of the grand canonical free energy density. In flat space and without external gauge fields, ${\cal F}$ reduces to the pressure $P$, and in general ${\cal F}$ is a complicated function of the spatially varying external sources. In a slight abuse of terminology, we will refer to ${\cal F}$ as the free energy density, and to $W$ as the free energy. Varying $W$ with respect to a time-independent source gives rise to a zero-frequency insertion in the Euclidean path integral of the operator coupled to the source. The relevant operators are the energy-momentum tensor (coupled to the metric), and the conserved current (coupled to the gauge field). Thus $W$ is the generating functional for zero-frequency correlation functions of the energy-momentum tensor and the current in equilibrium. 

The Euclidean external sources $\gE_{\mu\nu}$ and $\AE_\mu$ may be ``un-Wick-rotated'' to Minkowski time to obtain the physical real-time external sources $g_{\mu\nu}$ and $A_\mu$, for example $\gE_{00}=-g_{00}$, $\gE_{0k}=-ig_{0k}$, $\AE_0=-iA_0$ etc. In what follows we will omit the dependence on $T_0$ and $\mu_0$, and will denote the Euclidean generating functional with arguments continued to physical time as $W[g,A]$, so that
\begin{equation}
\label{eq:WM}
  W[g,A] = \int\!d^{d+1}x\; 
  \sqrt{-g}\, {\cal F}(T_0,\mu_0, g, A)\,,  
\end{equation}
where $\int d^{d+1}x$ stands for $-i\beta_0\int d^dx$. We may as well view $\int d^{d+1}x$ as containing an integral over the physical time, as the argument of the integral does not depend on time anyway. 

For a relativistic microscopic theory without gauge and gravitational anomalies, the generating functional is both gauge- and diffeomorphism-invariant. 
Let us further assume that all long-range interactions are screened due to a non-zero temperature $T$, so that the spatial correlations are local on scales longer than the screening length. The effective description of static correlations on such long scales will then be given by $W[g,A]$, where the density ${\cal F}$ is a local function of the external sources. For external fields that vary slowly in space, the above locality implies that ${\cal F}$ may be written as a derivative expansion in the gradients of the external fields. 
See Ref.~\cite{Banerjee:2012iz} for a study of the local generating functional in the Euclidean form, and Ref.~\cite{Jensen:2012jh} for the Minkowski form. Here we will use the Minkowski form, in which the underlying gauge and diffeomorphism invariance is manifested in a more straightforward way.

In order to implement the derivative expansion in practice, one needs to postulate the derivative scaling of the external sources $g_{\mu\nu}$ and $A_\mu$. Physically, this amounts to deciding whether the external sources are taken as ``strong'' or ``weak'' on the scale of the spatial inhomogeneity in equilibrium. In Refs.~\cite{Banerjee:2012iz, Jensen:2012jh}, the external sources were taken as ``weak'' in the sense that both $g_{\mu\nu}$ and $A_\mu$ were assumed to be $O(1)$ in the derivative expansion, so that both electric and magnetic fields appear at order $O(\partial)$ in the expansion. This choice of scaling makes the description of equilibrium polarization rather awkward: for example, in 3+1 dimensions, the thermodynamic response to constant homogeneous magnetic field ${\bf B}$ appears at the same order as the response to two derivatives of temperature, ${\bf B}^2\sim (\partial T)^2$.

In order to describe polarized matter in constant (or slowly varying in space) electric and magnetic fields, a different derivative counting scheme is more natural, one in which constant homogeneous electric and magnetic fields are taken to be $O(1)$ in the derivative expansion, rather than order $O(\partial)$.
This will be our goal here: to implement the derivative expansion of the free energy in the regime when the external gravitational field is still ``weak'' so that $g_{\mu\nu}$ is $O(1)$, while the external electromagnetic field is ``strong'' so that the  field strength $F_{\mu\nu}=\partial_\mu A_\nu - \partial_\nu A_\mu$ is $O(1)$. Following the general approach of Ref.~\cite{Jensen:2012jh}, we will obtain simple expressions for the energy-momentum tensor and the conserved current in relativistic polarized matter subject to external fields.

\section{Free energy}

\subsection{Thermodynamic parameters}
Let us first outline the starting point. The free energy $W[g,A]$ is a gauge- and diffeomorphism-invariant functional of $A_\mu$, $g_{\mu\nu}$, and their derivatives. Being in equilibrium means that that there is a Killing vector $V$, such that the Lie derivative with respect to $V$ vanishes on all observables, $\L_V(\dots) = 0$. In suitable coordinates, $V^\mu=(1,{\bf 0})$. In the grand canonical ensemble, the equilibrium state is parametrized by the temperature, velocity, and the chemical potential. Their relation to the external sources is
\begin{equation}
\label{eq:Tumu}
  T = \frac{1}{\beta_0 \sqrt{-V^2}}\,,\ \ \ \ 
  u^\mu = \frac{V^\mu}{\sqrt{-V^2}}\,,\ \ \ \ 
  \mu = \frac{ V^\mu A_\mu + \Lambda_V}{\sqrt{-V^2}}\,,
\end{equation}
where $\beta_0$ is a constant setting the normalization of temperature, and $\Lambda_V$ is a gauge parameter which ensures that $\mu$ is gauge-invariant. The constant $\mu_0$ is absorbed into $\Lambda_V$. Without external gauge fields, relations~(\ref{eq:Tumu}) are the covariant versions of the statement that $T\sqrt{-g_{00}}$ and $\mu \sqrt{-g_{00}}$ are constant in equilibrium~\cite{LL5}. The vector $u^\mu$ is the normalized ($u^2=-1$) velocity of matter, and the coordinates in which $V^\mu=(1,{\bf 0})$ correspond to the matter ``at rest''.  Both $T$ and $\mu$ are gauge invariant and transform as scalars under diffeomorphisms. For a discussion of gauge and diffeomorphism covariance of the equilibrium parameters see sec.~5 of Ref.~\cite{Jensen:2013kka}.
For a system occupying a spacetime region $\M$ with a boundary $\partial\M$, we assume that the generating functional can be separated into bulk and boundary contributions, and we take 
\begin{equation}
\label{eq:W0}
  W[g,A] = \int_{\M}\! d^{d+1}x\; \sqrt{-g}\,  {\cal F}
  + \int_{\partial\M} d^d x\; \sqrt{\gamma}\, L \,.
\end{equation}
Here the first term describes the bulk contribution, and the second term the boundary contribution. To leading order in the derivative expansion, ${\cal F}$ is the pressure, and $L$ is the surface tension. In the bulk term, $g$ is the determinant of $g_{\mu\nu}$, and ${\cal F}$ is a function of $T$, $u^\mu$, $\mu$, as well as of the sources $A_\mu$, $g_{\mu\nu}$, and their derivatives. For the boundary with coordinates $y^a$ whose shape is specified by $x^\mu(y^a)$, the tangent vectors are $e^\mu_a=\partial x^\mu/\partial y^a$, and the projector onto the boundary is $P^{\mu\nu}=g^{\mu\nu}-n^\mu n^\nu$. The induced metric on the boundary is $\gamma_{ab} = e^\mu_a e^\nu_b g_{\mu\nu}$. In the boundary term, $\gamma$ is the determinant of the induced metric, while $L$ in addition may depend on $n^\mu$, the spacelike unit normal vector to the boundary.

\subsection{Response to external sources}

The energy-momentum tensor and the current are defined as
\begin{align}
\label{eq:TJ}
  \delta W[g,A] & = \coeff12 \int_{\M}\!d^{d+1}x\; \sqrt{-g}\, T^{\mu\nu} \, \delta g_{\mu\nu}
  + \int_{\M}\!d^{d+1}x\; \sqrt{-g}\, J^{\mu}\, \delta A_{\mu} \nonumber\\[5pt]
  & + \coeff12 \int_{\partial\M}\!d^{d}x\; \sqrt{\gamma}\, T_\S^{\mu\nu} \, \delta g_{\mu\nu}
  + \int_{\partial\M}\!d^{d}x\; \sqrt{\gamma}\, J_\S^{\mu}\, \delta A_{\mu} \nonumber\\[5pt]
  & +\int_{\partial\M} \! d^dx\, \sqrt{\gamma}\, K^{\mu\nu}_\S\, \L_n \delta g_{\mu\nu}
    + \int_{\partial\M} \! d^dx\, \sqrt{\gamma}\, K^\mu_\S\, \L_n \delta A_\mu
  + \dots\,,
\end{align}
where $\L_n$ is the Lie derivative along the normal, the dots denote boundary terms with higher normal derivatives of the sources. The variations are performed at fixed $V^\mu$ and $\Lambda_V$. Here $T^{\mu\nu}$, $J^\mu$ are the bulk energy-momentum tensor and the current, and $T_\S^{\mu\nu}$, $J_\S^\mu$ are the boundary energy-momentum tensor and the current. The tensors $K^{\mu\nu}_\S$ and $K^{\mu}_\S$ describe interactions between bulk and boundary degrees of freedom. The gauge- and diffeomorphism-invariance of $W[g,A]$ implies conservation laws
\begin{subequations}
\label{eq:TJ-conserv0}
\begin{align}
  & \nabla_{\!\mu} J^\mu = 0\,,\\
  & \nabla_{\!\mu} T^{\mu\nu} = F^{\nu\lambda} J_\lambda\,.
\end{align}
\end{subequations}
The derivative expansion for the free energy density ${\cal F}$ in the generating functional (\ref{eq:W0}) gives rise to the derivative expansion for the equilibrium $T^{\mu\nu}$ and $J^\mu$, as described in~\cite{Jensen:2012jh}.
The boundary energy-momentum tensor and the current in (\ref{eq:TJ}) may be decomposed into the contributions tangential to and normal to the boundary,
\begin{align*}
  & T_\S^{\mu\nu} \, \delta g_{\mu\nu} = T_\S^{ab}\,\delta \gamma_{ab} + \Pi^a_\S\, \delta g_a + \Pi_\S\, \delta g_\n\,,\\
  & J_\S^\mu\, \delta A_\mu = J_\S^a \delta A_a + J_\S\, \delta A_{\n}\,.
\end{align*}
Here $\delta g_a$ is the pullback of $P_\mu^\lambda\, \delta g_{\lambda\nu} n^\nu$ to the boundary, $\delta g_\n = n^\mu n^\nu \delta g_{\mu\nu}$, $\delta A_a$ is the pullback of $\delta A_\mu$ to the boundary, and $\delta A_{\n} = n^\mu \delta A_{\mu}$.
Similarly, one can vary the generating functional with respect to the field strength $F_{\mu\nu}$,
\begin{align}
  \delta_F W[g,A] & = 
  \coeff12 \int_{\M}\!d^{d+1}x\; \sqrt{-g}\; M^{\mu\nu} \, \delta F_{\mu\nu}
  +\coeff12\int_{\partial\M}\!d^{d}x\; \sqrt{\gamma}\, M_\S^{ab}\,\delta F_{ab}
  \nonumber\\[5pt]
  & + \int_{\partial\M} \! d^dx\, \sqrt{\gamma}\,
    M^a_\S\, \delta F_a 
  +\int_{\partial\M} \! d^dx\, \sqrt{\gamma}\, L^{\mu\nu}_\S\, \L_n \delta F_{\mu\nu} + \dots\,
\label{eq:dFWS0}
\end{align}
where again $\delta F_{ab}$ is the pullback of $\delta F_{\mu\nu}$ to the boundary, $\delta F_a$ is the pullback of $P_\mu^\lambda\, \delta F_{\lambda\nu} n^\nu$ to the boundary,  and the dots denote boundary terms with higher normal derivatives of $\delta F_{\mu\nu}$. This defines the bulk polarization tensor $M^{\mu\nu}$, and the boundary polarization tensor $M_\S^{ab}$.%
\footnote{
There is a gravitational analogue of the polarization tensor which involves varying the generating functional with respect to the connection coefficients. The energy-momentum tensor then takes the form analogous to Eq.~(\ref{eq:JJ}) below. See sec.~5 of Ref.~\cite{Jensen:2013kka}.
} 
The surface terms $J_\S$, $\Pi^a_\S$, $\Pi_\S$, $M^a_\S$ depend on how the equilibrium is set up, and what the boundary conditions on $\partial \M$ are, as determined by the nature of the phase separation at~$\partial \M$. 

In all the above variations, we assume that the region $\M$ occupied by matter is unchanged. One could also consider the response of the generating functional to changing the shape of $\partial\M$, however this will not be needed for our purposes. See Ref.~\cite{Armas:2015ssd} for a recent discussion of surface terms in the Euclidean generating functional.

The polarization tensor contains both electric and magnetic components. We define the electric field as $E_\mu\equiv F_{\mu\nu}u^\nu$, the magnetic field as $B\equiv -\coeff12 \epsilon^{\mu\alpha\beta} u_\mu F_{\alpha\beta}$ for $d=2$, and $B^\mu \equiv \coeff12 \epsilon^{\mu\nu\alpha\beta} u_\nu F_{\alpha\beta}$ for $d=3$. 
In 1+1 dimensions, we define the ``magnetic field'' as $B\equiv \coeff12 \epsilon^{\mu\nu}F_{\mu\nu}$, so that $F_{\mu\nu} = -B\epsilon_{\mu\nu}$.
The Levi-Civita tensor is $\epsilon^{\mu\nu\alpha\beta}= \varepsilon^{\mu\nu\alpha\beta}/\sqrt{-g}$, with $\varepsilon^{0123}=1$, and similarly in other dimensions.  Both $E_\mu$ and $B_\mu$ are spacelike and orthogonal to $u^\mu$. 
We have the following decomposition of the field strength:
\begin{align*}
 & d=1:\ \ \ \   F_{\mu\nu} = u_\mu E_\nu - u_\nu E_\mu
                 \,,\\[5pt]
 & d=2: \ \ \ \  F_{\mu\nu} = u_\mu E_\nu - u_\nu E_\mu 
                 -\epsilon_{\mu\nu\rho} u^\rho B\,,\\[5pt]
 & d=3: \ \ \ \  F_{\mu\nu} = u_\mu E_\nu - u_\nu E_\mu 
                 -\epsilon_{\mu\nu\rho\sigma} u^\rho B^\sigma\,.
\end{align*}

The electric polarization vector $p^\alpha$ and the magnetization vector $m^\alpha$ (for $d{=}3$) are defined by rewriting the integrand in (\ref{eq:dFWS0}) as $\frac12 M^{\mu\nu} \delta F_{\mu\nu} = p^\alpha \delta E_\alpha + m^\alpha \delta B_\alpha$. For $d{=}2$, the variation is $\frac12 M^{\mu\nu} \delta F_{\mu\nu} = p^\alpha \delta E_\alpha + m \delta B$, which defines the magnetization $m$. The decomposition of the polarization tensor into the electric and magnetic parts is then
\begin{subequations}
\label{eq:MMG}
\begin{align}
 & d=1: \ \ \ \ M_{\mu\nu} = m \epsilon_{\mu\nu}\,,\\[5pt]
\label{eq:MMG-2}
 & d=2: \ \ \ \ M_{\mu\nu} = p_\mu u_\nu - p_\nu u_\mu 
                 -\epsilon_{\mu\nu\rho} u^\rho m\,,\\[5pt]
 & d=3: \ \ \ \  M_{\mu\nu} = p_\mu u_\nu - p_\nu u_\mu 
                 -\epsilon_{\mu\nu\rho\sigma} u^\rho m^\sigma\,,
\end{align}
\end{subequations}
where $p^\mu \equiv u_\lambda M^{\lambda\mu}$, $m^\mu \equiv \coeff12 \epsilon^{\mu\nu\alpha\beta} u_\nu M_{\alpha\beta}$ (for $d=3$), and $m \equiv -\coeff12 \epsilon^{\mu\alpha\beta} u_\mu M_{\alpha\beta}$ (for $d=2$). Both $p_\alpha$ and $m_\alpha$ are transverse to $u^\alpha$.

\subsection{Equilibrium relations}
The equilibrium definitions (\ref{eq:Tumu}) together with $\L_V(\dots)=0$ give
\begin{equation}
\label{eq:Tmueq}
  \partial_\lambda T = - T a_\lambda\,,\ \ \ \ 
  \partial_\lambda \mu = E_\lambda - \mu a_\lambda \,,
\end{equation}
where $a^\mu \equiv u^\lambda \nabla_{\!\lambda}u^\mu$ is the acceleration vector, $u_\mu a^\mu=0$.
These relations imply that $T\partial_\lambda(\mu/T) - E_\lambda$ vanishes in equilibrium. In other words, a system subject to an external electric field will develop a gradient of $\mu/T$ in order to compensate the applied field and ensure that the equilibrium is maintained.

This has implication for derivative counting. For ``weak'' electric fields $E_\lambda \sim O(\partial)$, the gradients of $T$ and $\mu$ are $O(\partial)$ as well. For ``strong'' electric fields $E_\lambda \sim O(1)$, there will be an $O(1)$ gradient of $\mu/T$. How exactly this gradient is achieved depends on the nature of the microscopic degrees of freedom. Given that the chemical potential determines the number of charge carriers, we take ``strong'' electric fields to mean that both $E$ and $\partial\mu$ are $O(1)$, while $\partial T$ is still $O(\partial)$, so that $\frac{\partial\mu}{\mu}\gg \frac{\partial T}{T}$. In the generating functional, the derivatives of the chemical potential may then be traded for the electric field.

Similarly, the derivative of the velocity can be decomposed in equilibrium as 
\begin{subequations}
\label{eq:dueq}
\begin{align}
  & d=1:\ \ \ \ \nabla_{\!\mu} u_\nu = -u_{\mu} a_\nu\,,\\[5pt]
  & d=2:\ \ \ \ \nabla_{\!\mu} u_\nu = -u_{\mu} a_\nu 
                -\coeff12 \epsilon_{\mu\nu\alpha}\, u^\alpha\Omega\,,\\[5pt]
  & d=3:\ \ \ \ \nabla_{\!\mu} u_\nu = -u_{\mu} a_\nu 
                -\coeff12 \epsilon_{\mu\nu\alpha\beta}\, u^\alpha\Omega^\beta \,,
\end{align}
\end{subequations}%
The vorticity is $\Omega \equiv -\epsilon^{\mu\nu\lambda}u_\mu \nabla_{\!\nu} u_\lambda$ for $d=2$, and $\Omega^\mu \equiv \epsilon^{\mu\nu\alpha\beta} u_\nu \nabla_{\!\alpha} u_\beta$ for $d=3$. This velocity decomposition implies that both the expansion $\nabla_{\!\mu}u^\mu$ and the shear tensor $\sigma^{\mu\nu}\equiv (\Delta^{\mu\alpha} \Delta^{\nu\beta} + \Delta^{\nu\alpha}\Delta^{\mu\beta} - \frac2d \Delta^{\mu\nu} \Delta^{\alpha\beta})\nabla_{\!\alpha} u_\beta$ (where $\Delta^{\mu\nu} = g^{\mu\nu} + u^\mu u^\nu$ is the transverse projector) vanish in equilibrium. This is as it should be: out of equilibrium, the expansion would contribute to dissipation through bulk viscosity, and the shear tensor would contribute to dissipation through shear viscosity. 

Combined with the electromagnetic ``Bianchi identity'' $\epsilon^{\mu\nu\alpha\beta}\nabla_{\!\nu}F_{\alpha\beta}=0$ in 3+1 dimensions, the velocity decomposition (\ref{eq:dueq}) implies
\begin{align*}
  & \nabla{\cdot}B = B{\cdot}a - E{\cdot}\Omega\,,\\[5pt]
  & u_\mu \epsilon^{\mu\nu\rho\sigma}\nabla_{\!\rho} E_\sigma = u_\mu \epsilon^{\mu\nu\rho\sigma}E_\rho a_\sigma\,.
\end{align*}
These are the covariant versions of the familiar flat-space equilibrium relations ${\bm \nabla}{\cdot}{\bf B}=0$ and ${\bm \nabla}{\times} {\bf E}=0$.
More generally, for the electric field in equilibrium we have 
\begin{subequations}
\begin{align}
  & d=2:\ \ \ \ \epsilon^{\mu\alpha\beta}\nabla_{\!\alpha} E_\beta 
    = \epsilon^{\mu\alpha\beta}E_\alpha a_\beta\,,\\[5pt]
  & d=3:\ \ \ \ \epsilon^{\mu\nu\rho\sigma}\nabla_{\!\rho} E_\sigma 
    = \epsilon^{\mu\nu\rho\sigma}E_\rho a_\sigma \,,
\end{align}
\end{subequations}
as a consequence of $\L_V E_\alpha =0$ and $E_\alpha u^\alpha=0$.

\subsection{Polarization ambiguities}
\label{sec:ambiguities}
The electromagnetic Bianchi identity also implies that there is an ambiguity in the definition of the polarization tensor: in 3+1 dimensions, one can always add to the generating functional an identically vanishing term $W_{\!\varnothing}=\frac12\int\! \sqrt{-g}\,C_\mu\, \epsilon^{\mu\nu\alpha\beta}\nabla_{\!\nu}F_{\alpha\beta}$, where $C_\mu$ can be a function of the field strength and its derivatives. Such a term shifts the polarization tensor by
\begin{equation}
\label{eq:M_3}
  M^{\alpha\beta} \to M^{\alpha\beta} 
  + \epsilon^{\alpha\beta\mu\nu} \nabla_{\!\mu} C_\nu \,.
\end{equation}
The polarization vectors correspondingly shift as
\begin{subequations}
\label{eq:pm-shift}
\begin{align}
  & p^\mu \to p^\mu - \epsilon^{\mu\nu\rho\sigma}u_\nu \nabla_{\!\rho}C_\sigma\,,\\[5pt]
  & m_\mu \to m_\mu - \nabla_{\!\mu}(C{\cdot}u) - (C{\cdot}u)a_\mu\,.
\end{align}
\end{subequations}
In 2+1 dimensions, we can add an identically vanishing term $W_{\!\varnothing} = \frac12 \int \! \sqrt{-g}\,C \epsilon^{\mu\alpha\beta}\nabla_{\!\mu}F_{\alpha\beta}$, where again $C$ can be a function of the field strength and its derivatives. The polarization tensor then shifts by
\begin{equation}
\label{eq:M_2}
  M^{\alpha\beta} \to M^{\alpha\beta} 
  - \epsilon^{\alpha\beta\mu}\, \nabla_{\!\mu} C \,.
\end{equation}
The electric polarization vector correspondingly changes as
\begin{align}
\label{eq:p2-amb}
  p^\mu \to p^\mu + \epsilon^{\mu\nu\rho}u_\nu \nabla_{\!\rho}C  \,,
\end{align}
while the magnetic polarization $m$ remains unchanged. 

The variational derivatives of $W_{\!\varnothing}$ with respect to both $g_{\mu\nu}$ and $A_\mu$ vanish. As a result, the energy-momentum tensor and the current (both bulk and boundary) are not affected by such unphysical shifts.

\subsection{Bound charges and bound currents}
For matter whose degrees of freedom carry gauge charges, it is conventional to separate the charge into the ``free charge'' and ``bound charge'' components. In the grand canonical ensemble, the chemical potential $\mu$ describes the coupling of the system to a reservoir of ``free charges''. Demanding local charge neutrality for free charges in the bulk would amount to demanding $\partial {\cal F}/\partial\mu=0$. Doing so would eliminate the contribution of free charges to polarization. One may refer to $\mu$-dependent contributions as coming from ``free charges'', and $\mu$-independent contributions as coming from ``bound charges'', though such a separation is somewhat artificial. We will not impose $\partial {\cal F}/\partial\mu=0$, and will keep the contribution to polarization from both free charges and bound charges. 

The current $J^\mu$ admits a simple expression in terms of the polarization tensor to any order in the derivative expansion. Indeed, the free energy density ${\cal F}$ can be written as
$$
  {\cal F} = S_{(0)} + S_{(1)}^{\lambda\mu\nu}\nabla_{\!\lambda}F_{\mu\nu}
  + S_{(2)}^{\lambda\sigma\mu\nu}\nabla_{\!\lambda}\nabla_{\!\sigma}F_{\mu\nu}
  +\dots\,,
$$
where the coefficients $S_{(n)}$ do not contain derivatives of the electromagnetic field strength. The derivative of the chemical potential can be traded for the electric field according to (\ref{eq:Tmueq}), hence we can take $S_{(n)}=S_{(n)}(T,\mu,F_{\alpha\beta},\dots)$ where dots denote the arguments which do not depend on the gauge field. The polarization tensor can be easily found in terms of $S_{(n)}$ through integration by parts. It is then clear that the current extracted from the generating functional according to (\ref{eq:TJ}) is 
\begin{equation}
\label{eq:JJ}
  J^\alpha = \rho u^\alpha  
  - \nabla_{\!\lambda} M^{\lambda\alpha}\,,
\end{equation}
to any order in the derivative expansion, where $\rho\equiv\partial{\cal F}/\partial\mu$. The first term in the right-hand side is the standard equilibrium current in the absence of polarization: to leading order in the derivative expansion the free energy density ${\cal F}$ is just the pressure $P$, and $\rho=\partial P/\partial\mu$  is the  density of ``free charges''. 
The second term in the right-hand side is a total derivative of an anti-symmetric tensor. It therefore does not contribute to the conservation equation $\nabla_{\!\mu} J^\mu = 0$, and can be interpreted in terms of ``bound'' charges and ``bound'' currents. It is clear from the expression (\ref{eq:JJ}) that the unphysical polarization shifts (\ref{eq:M_3}) and (\ref{eq:M_2}) do not affect the current.

The current can be decomposed with respect to the velocity $u^\mu$ as
\begin{equation}
  J^\mu = {\cal N} u^\mu + {\cal J}^\mu\,,
\end{equation}
where ${\cal N}\equiv -u_\mu J^\mu$ is the charge density, and the spatial current ${\cal J}_\mu \equiv \Delta_{\mu\lambda} J^\lambda$ is transverse to~$u^\mu$. For the polarization tensor of the form (\ref{eq:MMG}), the definitions (\ref{eq:Tumu}) together with $\L_V(\dots)=0$ lead to the following equilibrium expressions for the charge density:
\begin{subequations}
\label{eq:NN23}
\begin{align}
  & d=2: \ \ \ \ {\cal N} = \rho - \nabla_{\!\mu}\, p^\mu + p^\mu a_\mu -m \Omega\,,\\[5pt]
  & d=3: \ \ \ \ {\cal N} = \rho - \nabla_{\!\mu}\, p^\mu + p^\mu a_\mu -m_\mu \Omega^\mu\,.
\end{align}
\end{subequations}
Consider the charge density in $d=3$ spatial dimensions. The second term in the right-hand side is the familiar electrostatic bound charge density, which in flat space reduces to $-\grad\!{\cdot}\p$. The third term is the bound charge density induced by gravity: in the static Newtonian gravitational field it becomes ${\bf p}{\cdot}\! \grad\varphi$, where  $\varphi$ is the gravitational potential. The last term is the bound charge density induced in  magnetized matter which is rotating. For a system undergoing rotation with small (meaning $|{\bm \omega}|R\ll 1$, where $R$ is the size of the system) angular velocity ${\bm \omega}$, the last term in the right-hand side becomes $-2{\bf m}{\cdot}{\bm\omega}$.

Similarly, the definitions (\ref{eq:Tumu}) together with $\L_V=0$ lead to the following equilibrium expressions for the spatial current:
\begin{subequations}
\label{eq:JJ23}
\begin{align}
  & d=2:\ \ \ \ 
  {\cal J}^\mu = \epsilon^{\mu\nu\rho} u_\nu \nabla_{\!\rho} m 
    + \epsilon^{\mu\nu\rho} u_\nu a_\rho m \,,\\[5pt]
  & d=3:\ \ \ \ 
  {\cal J}^\mu = \epsilon^{\mu\nu\rho\sigma} u_\nu \nabla_{\!\rho} m_\sigma
  +\epsilon^{\mu\nu\rho\sigma} u_\nu a_\rho m_\sigma \,.
\end{align}
\end{subequations}
Consider the current density in $d=3$ spatial dimensions. The first term in the right-hand side is the familiar bound current, which in flat space reduces to $\grad{\times}{\bf m}$. The second term is the bound current induced by the gravitational field: in the static Newtonian gravitational field it reduces to $(\grad\varphi){\times}{\bf m}$, where  $\varphi$ is the gravitational potential. 

We emphasize that the above expressions for bound charges and bound currents are simply a consequence of thermal equilibrium. Equations (\ref{eq:NN23}), (\ref{eq:JJ23}) do not assume any particular microscopic model of matter, and moreover they hold to any order in the derivative expansion.

\subsection{Derivative expansion}
\label{sec:derivative-expansion}
We close this section with a comment on the derivative expansion of the free energy. As a schematic example, consider the functional $W[a,g]$ which depends on two sources $a(x)$ and $g(x)$ which both vary slowly in space. Assuming locality, the derivative expansion is
\begin{align*}
  W[a,g] & = \int\! P_0(a,g) + \int\! P_1(a,g)\, a'(x) + \int\! P_2(a,g)\, g'(x) \\[5pt]
  & + \int\! P_3(a,g)\, a''(x) + \int\! P_4(a,g)\, g''(x) 
    + \int\! P_5(a,g)\, a'(x) g'(x) \\[5pt] 
  & + \int\! P_6(a,g)\, a'(x)^2 
    + \int\! P_7(a,g)\, g'(x)^2 + O(\partial^3) \,,
\end{align*}%
and the boundary terms are implied. 

Suppose now that $a$ changes much faster than $g$, such that $|a'(x)/a(x)| \gg |g'(x)/g(x)|$. Naively, one may think that the terms containing the derivatives of $a(x)$ are more important than those with derivatives of $g(x)$, and there is a separate derivative counting associated with $a(x)$ and $g(x)$. This is not in general so: for example, integrating the $P_2$ term by parts gives rise to $\int (\partial P_2/\partial a)\, g(x) a'(x)$, which may be of the same order as the $P_1$ term. It is possible to count the derivatives of $g$ differently from the derivatives of $a$ if the ``cross'' terms $(\partial P_2/\partial a)$ are in some sense small. For example, we could introduce two counting parameters $\varepsilon$ and $\gamma\ll \varepsilon$ and count the derivatives as $a'\sim \varepsilon$, $g'\sim\gamma$, while $\partial P_2/\partial a$, $\partial P_4/\partial a$, $\partial P_5/\partial a$, $\partial P_7/\partial a$ are of order $\gamma/\varepsilon$.

Physically, $a$ will be the external gauge potential $A_\mu$, and $g$ the external metric $g_{\mu\nu}$. By ``strong'' electromagnetic fields we will mean the fields such that this derivative counting is valid, i.e.\ electromagnetism is more important than gravity. For such ``strong'' fields, by the leading order in the derivative expansion we will mean: {\em i)} setting $\gamma$ to zero, {\em ii)} isolating terms polynomial in $a'(x)$, and {\em iii)} summing those terms into a single function $P(a,g,a')$.

\section{Leading order in the derivative expansion}

\subsection{Weak electromagnetic fields}
Let us start with ``weak'' electromagnetic fields. For the sources with $A_\mu\sim O(1)$, $g_{\mu\nu}\sim O(1)$, there are only two gauge and diffeomorphism invariants at leading order in the derivative expansion in the bulk: $T$ and $\mu$. On the boundary, there is an extra invariant $u_n \equiv u^\mu n_\mu$. The static generating functional to leading order in the derivative expansion is then
$$
  W[g,A] = \int_{\M}\!d^{d+1}x\; \sqrt{-g}\, P(T,\mu)
  + \int_{\partial \M}\!d^{d}x\; \sqrt{\gamma}\, L(T,\mu,u_n)\,.
$$
The definitions (\ref{eq:TJ}) give
\begin{align*}
  & T^{\mu\nu} = P g^{\mu\nu} + (Ts + \mu \rho) u^\mu u^\nu\,,\\
  & J^\mu = \rho u^\mu\,,
\end{align*}
where $s = \partial P/\partial T$ is the bulk entropy density, $\rho = \partial P/\partial\mu$ is the bulk charge density. These are the standard expressions for the energy-momentum tensor and the current in a relativistic perfect fluid. The boundary energy-momentum tensor and the current are
\begin{align*}
  & T^{ab}_\S = L\gamma^{ab} + 
    \left( T s_\S +\mu \rho_\S \right) u^a u^b\,,\\[5pt]
  & J^a_\S = \rho_\S u^a\,,
\end{align*}
and the other boundary terms are $\Pi^a_\S = \left( T s_\S + \mu \rho_\S \right) u_n u^a$, $\Pi_\S = \coeff12 \left( T s_\S+ \mu \rho_\S\right)u_n^2 +\varepsilon_\S u_n$, and $J_\S = \rho_\S u_n$, where we have defined $s_\S \equiv {\partial L}/{\partial T}$, $\rho_\S \equiv {\partial L}/{\partial \mu}$, $\varepsilon_\S \equiv \partial L/\partial u_n$. Again, these describe a perfect fluid on the boundary with pressure~$L$. At leading order in the derivative expansion, both $K^\mu_\S$ and $K^{\mu\nu}_\S$ vanish.

\subsection{Strong electromagnetic fields}
Let us now consider ``strong'' electromagnetic fields, such that $F_{\mu\nu}\sim O(1)$ and $g_{\mu\nu}\sim O(1)$ in the derivative expansion. To leading order, the static generating functional is
\begin{equation}
\label{eq:W00}
  W[g,A] = \int_{\M}\!d^{d+1}x\; \sqrt{-g}\, P(T, \mu, F_{\alpha\beta})
  + \int_{\partial \M}\!d^{d}x\; \sqrt{\gamma}\, L(T, \mu, F_{\alpha\beta}, n_\alpha)\,.
\end{equation}
The dependence on $F_{\alpha\beta}$ includes the dependence on electric and magnetic fields, and for the boundary part also on their normal components.%
\footnote{
When $d{+}1$ is odd, there may be Chern-Simons terms in $P$. The Chern-Simons term is not gauge invariant on the boundary, so in this case $L$ must contain an anomalous piece, whose gauge variation exactly cancels the gauge variation of the Chern-Simons term. In the application to the quantum Hall effect, the anomalous boundary piece comes from the massless 1+1 dimensional chiral modes on the boundary~\cite{Wen-FQHE}. Upon integrating out the massless boundary modes, $L$ will in general become a non-local function of the electric and magnetic fields. If the dynamics of the boundary modes can be described classically, they may be treated directly within the generating functional, similar to what is done in Ref.~\cite{Bhattacharyya:2012xi} for superfluids. In what follows, we will ignore the massless boundary modes, and will only explore the consequences of short-distance correlations on the boundary.
}
 
The bulk current is given by (\ref{eq:JJ}), with the polarization tensor $M^{\mu\nu} = 2\partial P/\partial F_{\mu\nu}$. In what follows we will express $M^{\mu\nu}$ in terms of electric and magnetic susceptibilities.

In order to find the energy-momentum tensor, we need to be more specific about the dependence of $P$ and $L$ on the metric. Gauge and diffeomorphism invariance requires that $P=P(s^{\scriptscriptstyle (0)})$ is a function of scalars $s^{\scriptscriptstyle (0)}$, which are made out of the electromagnetic field strength (we will use the term ``scalar'' for both scalars and pseudo-scalars). The superscript signifies that we are working to leading order in the derivative expansion. 
The number of scalars $s^{\scriptscriptstyle (0)}$ depends on the dimension. To leading order in the derivative expansion, we choose to work with the following independent scalars:
\begin{subequations}
\label{eq:scalars-0}
\begin{align}
   d=1: \ \ & s^{\scriptscriptstyle (0)} = \left\{T,\ \mu,\ B \right\}\,,\\[5pt]
   d=2: \ \ & s^{\scriptscriptstyle (0)} = \left\{T,\ \mu,\ B,\ E^2 \right\}\,,\\[5pt]
   d=3: \ \ & s^{\scriptscriptstyle (0)} = \left\{T,\ \mu,\ B^2,\ E{\cdot}B,\ E^2 \right\}\,.
\end{align} 
\end{subequations}
Let us express the bulk energy-momentum tensor using the decomposition with respect to the velocity $u^\mu$, as is often done in relativistic fluid dynamics,
\begin{equation}
\label{eq:Tmn0}
  T^{\mu\nu} = {\cal E} u^\mu u^\nu + {\cal P}\Delta^{\mu\nu}
  + {\cal Q}^\mu u^\nu + {\cal Q}^\nu u^\mu + {\cal T}^{\mu\nu}\,.
\end{equation}
Here 
${\cal E}\equiv u_\mu u_\nu T^{\mu\nu}$ is the energy density, ${\cal P}\equiv \coeff{1}{d} \Delta_{\mu\nu}T^{\mu\nu}$ is the pressure, the momentum density ${\cal Q}_\mu \equiv -\Delta_{\mu\alpha} u_\beta T^{\alpha\beta}$ is transverse to $u^\mu$, and ${\cal T}^{\mu\nu}\equiv \coeff12(\Delta_{\mu\alpha}\Delta_{\nu\beta} + \Delta_{\nu\alpha} \Delta_{\mu\beta} - \coeff{2}{d} \Delta_{\mu\nu} \Delta_{\alpha\beta}) T^{\alpha\beta}$ is transverse to $u^\mu$, symmetric, and traceless. Given $P$ as a function of the above scalars, the energy-momentum tensor can be read off from the definition~(\ref{eq:TJ}).

For $d=1$, we have $P=P(T,\mu,B)$. This gives the following energy-momentum tensor:
\begin{subequations}
\label{eq:Tmn-1}
\begin{align}
  & {\cal E} = -P + Ts + \mu \rho + mB\,,\\
  & {\cal P} = P-mB\,,\\
  & {\cal Q}^\mu = 0\,,\\
  & {\cal T}^{\mu\nu} = 0\,.
\end{align}
\end{subequations}
Here the ``magnetization'' density $m\equiv \partial P/\partial B$ determines the polarization tensor as
\begin{equation}
\label{eq:MM-1}
  M^{\mu\nu} = m \epsilon^{\mu\nu}\,.
\end{equation}

For $d=2$,  $P=P(T,\mu, B, E^2)$. This gives the following energy-momentum tensor:
\begin{subequations}
\label{eq:Tmn-2}
\begin{align}
  & {\cal E} = -P + T s 
    + \mu \rho 
    + \aE E^2 
    \,,\\[5pt]
  & {\cal P} = P - m B  
    - \coeff12 \aE E^2 \,,\\[5pt]
  & {\cal Q}^\mu = m\, 
    \epsilon^{\mu\rho\sigma} E_\rho\, u_\sigma\,,\\[5pt]
  & {\cal T}^{\mu\nu} = -\aE
    \left(  E^\mu E^\nu - \coeff12 \Delta^{\mu\nu} E^2 \right)\,.
\end{align}
\end{subequations}
Here again $s = \partial P/\partial T$ is the entropy density, $\rho = \partial P/\partial\mu$ is the charge density, $m = \partial P/\partial B$ is the magnetization density, and $\aE \equiv 2\partial P/\partial E^2$ is the electric susceptibility. They determine the polarization tensor as
\begin{equation}
\label{eq:MM-2}
  M^{\mu\nu} = -m\,
  u_\lambda \epsilon^{\lambda\mu\nu}
  +\aE \left( E^\mu u^\nu - E^\nu u^\mu\right)\,.
\end{equation}
The dependence $P=P(T,\mu, B, E^2)$ implies that the electric polarization vector is $p^\mu = \aE E^\mu$, and the polarization tensor (\ref{eq:MM-2}) coincides with the general expression (\ref{eq:MMG}), as it should. 

For $d=3$, $P=P(T,\mu, B^2, E{\cdot}B, E^2)$. This gives the following energy-momentum tensor:%
\begin{subequations}
\label{eq:Tmn-3}
\begin{align}
  & {\cal E} = -P + T s 
    + \mu \rho 
    + \aEE E^2 + \aEB\, E{\cdot}B
    \,,\\[5pt]
  & {\cal P} = P - \coeff23 \aBB B^2  
    - \coeff13 \aEE E^2 -\aEB\, E{\cdot}B \,,\\[5pt]
  & {\cal Q}^\mu = -\aBB S^\mu \,,\\[5pt]
  & {\cal T}^{\mu\nu} = -\aEE
    \left(  E^\mu E^\nu - \coeff13 \Delta^{\mu\nu} E^2 \right)
    +\aBB \left(  B^\mu B^\nu - \coeff13 \Delta^{\mu\nu} B^2 \right)\,.
\end{align}
\end{subequations}
Here again $s = \partial P/\partial T$, $\rho = \partial P/\partial\mu$ are the entropy and charge densities, $S^\mu = \epsilon^{\mu\rho\sigma\lambda}u_\rho E_\sigma B_\lambda$ is the Poynting vector, $\aEE \equiv 2\partial P/\partial E^2$ is the electric susceptibility, $\aEB\equiv \partial P/\partial(E{\cdot}B)$ is the electro-magnetic susceptibility, $\aBB \equiv 2\partial P/\partial B^2$ is the magnetic susceptibility. They determine the polarization tensor as
\begin{equation}
\label{eq:MM-3}
  M^{\mu\nu} = \aEE (E^\mu u^\nu - E^\nu u^\mu)
  +\aBB \epsilon^{\mu\nu\alpha\beta} B_\alpha u_\beta
  -\aEB G^{\mu\nu}\,,
\end{equation}
where  $G^{\mu\nu} = \coeff12 \epsilon^{\mu\nu\alpha\beta}F_{\alpha\beta}$ is the dual field strength. 
The dependence $P=P(T,\mu, B^2, E{\cdot}B, E^2)$, implies that the polarization vectors are
\begin{align*}
  & p^\mu = \aEE E^\mu + \aEB B^\mu\,,\\
  & m^\mu = \aBE E^\mu + \aBB B^\mu\,.
\end{align*}
The magneto-electric susceptibility $\aBE$ is equal to the electro-magnetic susceptibility $\aEB$, and the polarization tensor (\ref{eq:MM-3}) coincides with the general expression (\ref{eq:MMG}), as it should. 

So far we have presented $T^{\mu\nu}$ in terms of the decomposition (\ref{eq:Tmn0}) with respect to the velocity $u^\mu$, whose coefficients ${\cal E}$, ${\cal P}$, ${\cal Q}^\mu$, and ${\cal T}^{\mu\nu}$ are expressed in terms of the electric and magnetic fields, and the susceptibilities. The same energy-momentum tensors (\ref{eq:Tmn-1}), (\ref{eq:Tmn-2}), (\ref{eq:Tmn-3}) can be equivalently expressed in terms of the polarization tensor $M^{\mu\nu}=2\partial P/\partial F_{\mu\nu}$ as
\begin{subequations}
\label{eq:Tmn-EM00}
\begin{equation}
\label{eq:Tmn-EM0}
  T^{\mu\nu} = P g^{\mu\nu} + (Ts + \mu  \rho) u^\mu u^\nu
  + T^{\mu\nu}_\EM\,,\\
\end{equation}
where
\begin{equation}
\label{eq:Tmn-EM}
  T^{\mu\nu}_\EM = M^{\mu\alpha} g_{\alpha\beta} F^{\beta\nu} 
  + u^\mu u^\alpha \left(
  M_{\alpha\beta} F^{\beta\nu} - F_{\alpha\beta} M^{\beta\nu} \right)
\end{equation}
\end{subequations}
is the ``electromagnetic correction'' to the perfect fluid form. Note that $P$, $s$, and $\rho$ in (\ref{eq:Tmn-EM0}) are functions of the electric and magnetic fields. The tensor (\ref{eq:Tmn-EM}) is symmetric; if we set the external electric field to zero (in two or three spatial dimensions), then $T^{\mu\nu}_\EM$ reduces to its first term, and is still symmetric. 
Note that $\nabla_{\!\mu}T^{\mu\nu}_\EM$ does not equal 
$F^\nu_{\ \;\lambda} (- \nabla_{\!\nu} M^{\nu\lambda})$.

The above expression for $T^{\mu\nu}_\EM$ was first derived by W.~Israel~\cite{Israel:1978up}, for a free gas of polarized relativistic particles.
We emphasize that one does not need to assume any particular microscopic model of matter in order to arrive at the above energy-momentum tensor: expression (\ref{eq:Tmn-EM00}) is a direct consequence of gauge and diffeomorphism invariance of the theory, to leading order in the derivative expansion.

We now turn to the boundary energy-momentum tensor and the current which follow from the generating functional~(\ref{eq:W00}). The boundary current may be expressed in terms of the boundary polarization tensor $m^{\mu\nu} \equiv 2\partial L/\partial F_{\mu\nu}$ (keeping $T$, $\mu$, and $n^\mu$ fixed). Upon integrating by parts on the boundary, the definition (\ref{eq:TJ}) gives the following boundary currents: 
\begin{align}
   J^a_\S = M^a + (\rho_\S u^a -\nabla_{\!b} m^{ba})\,,
\end{align}
Here $M^a$ is the boundary current arising from integrating the variation of $P(T, \mu, F_{\alpha\beta})$ by parts, $e^\mu_a M^a = n_\lambda M^{\lambda\mu}$. As one can see from the polarization tensor (\ref{eq:MMG}), in flat space in 3+1 dimensions the boundary current $n_\lambda M^{\lambda\mu}$ reduces to a vector whose time component (surface charge density) is $\p{\cdot}\n$, while the spatial part (surface bound current) is ${\bf m}{\times}\n$. These are the familiar expressions from electro- and magneto-statics. The other term in the boundary current, $(\rho_\S u^a -\nabla_{\!b} m^{ba})$, arises due to the presence of charged degrees of freedom on the boundary described by $L$, and mimics the bulk current (\ref{eq:JJ}), with $\rho_\S \equiv \partial L/\partial\mu$.
The other boundary currents, $J_\S = \rho_\S u_n$, and $K^\mu_\S = n_\lambda m^{\lambda \mu}$, emerge from $L$ as well.

At leading order in the derivative expansion, the only contribution to the boundary energy-momentum tensor arises from the surface tension term in~(\ref{eq:W00}). In 2+1 dimensions, $L=L(T,\mu, u_n, B, E^2, E_n)$, where $u_n \equiv u^\mu n_\mu$, $E_n\equiv E_\mu n^\mu$. At leading order $K^{\mu\nu}_\S$ vanishes, and the definition (\ref{eq:TJ}) gives
\begin{align*}
   T^{ab}_\S &  =  
   \left( T s_\S + \mu \rho_\S -m_\S B + \alpha_{\S,{\rm E}} E^2 
          +\chi_{n,{\rm E}} E_n \right) u^a u^b
   \nonumber\\[5pt]
   & + (L-m_\S B)\gamma^{ab} + m_\S (u^a S^b + u^b S^a) - \alpha_{\S,{\rm E}} E^a E^b\,.
\end{align*}
Here $S_a = e^\alpha_a P_{\alpha\mu} \epsilon^{\mu\rho\sigma}E_\rho u_\sigma$, and the coefficients are $s_\S \equiv {\partial L}/{\partial T}$, $\rho_\S \equiv {\partial L}/{\partial \mu}$, $m_\S\equiv \partial L/\partial B$, $\varepsilon_\S \equiv \partial L/\partial u_n$, $\alpha_{\S,{\rm E}}\equiv 2\partial L/\partial E^2$, $\chi_{n,{\rm E}}\equiv \partial L/\partial E_n$. The boundary energy-momentum tensor in 3+1 dimensions looks similar, and we won't write it down explicitly.

\section{Next order in the derivative expansion}
\label{sec:next-order}
We now proceed to the next (first) order in the derivative expansion, taking into account $O(\partial)$ terms in the generating functional~(\ref{eq:W0}). We will take the surface tension $L$ to be constant for simplicity, and will focus on the bulk contributions to thermodynamics. The free energy density at first order in the derivative expansion is 
\begin{equation}
\label{eq:F1}
  {\cal F} = P(s^{\scriptscriptstyle (0)}) 
  + \sum_n M_n(s^{\scriptscriptstyle (0)})\, s^{\scriptscriptstyle (1)}_n\,,
\end{equation}
where $P$ is the leading-order pressure. For weak electromagnetic fields, the leading order scalars are $s^{\scriptscriptstyle (0)}=\{T,\mu\}$, while for strong electromagnetic fields $s^{\scriptscriptstyle (0)}$ are given by Eq.~(\ref{eq:scalars-0}). The functions $M_n(s^{\scriptscriptstyle (0)})$ parametrize the thermodynamic response at first order, and are determined by the microscopic theory. 
The gauge- and diffeomorphism-invariant scalars $s^{\scriptscriptstyle (1)}_n$ depend on $T$, $u^\mu$, $\mu$, and the sources $A_\mu$, $g_{\mu\nu}$. The number of such first-order scalars depends on the dimension, and on whether the external electromagnetic fields are weak or strong. We will enumerate the scalars $s^{\scriptscriptstyle (1)}_n$ in what follows. 

The bulk current is still given by the general expression (\ref{eq:JJ}).
At first order, the free energy density (\ref{eq:F1}) may be equivalently rewritten as
\begin{equation}
\label{eq:Slmn}
  {\cal F} = P(s^{\scriptscriptstyle (0)}) 
  + P^{\prime}(s^{\scriptscriptstyle (0)}, s^{\scriptscriptstyle (1)})
  + \coeff12 S^{\lambda\mu\nu}(s^{\scriptscriptstyle (0)})
  \nabla_{\!\lambda} F_{\mu\nu}\,,
\end{equation}
where $P^{\prime}$ is $O(\partial)$, but contains no derivatives of $F_{\alpha\beta}$, while the last term parametrizes the static response to inhomogeneous electromagnetic fields. The polarization tensor is then
\begin{equation}
\label{eq:M2}
  M^{\alpha\beta} = 
  2\frac{\partial P}{\partial F_{\alpha\beta}} + 
  2\frac{\partial P^{\prime}}{\partial F_{\alpha\beta}}
  + \frac{\partial S^{\lambda\mu\nu}}{\partial F_{\alpha\beta}}
  \nabla_{\!\lambda}F_{\mu\nu}
  -\nabla_{\!\lambda}S^{\lambda\alpha\beta} \,.
\end{equation}
The equilibrium relation $\partial_\lambda \mu = E_\lambda +O(\partial)$ now implies that for strong electric fields the leading-order polarization tensor may receive contributions from subleading terms in the generating functional
$$
  M^{\alpha\beta} = 
  2\frac{\partial P}{\partial F_{\alpha\beta}} 
  - E_\lambda \frac{\partial S^{\lambda\alpha\beta}}{\partial\mu}
  + O(\partial)\,.
$$
The second term in the right-hand side describes a contribution of free charges to polarization. 
We will assume for simplicity that the effects of free charges are less important than those of bound charges, in the sense that $\partial M_n/\partial\mu \sim O(\partial)$. (Alternatively, the effects of the free charges may be lumped into the leading-order free energy, but in this case isolating their contribution becomes less straightforward.)
The boundary currents are
\begin{subequations}
\label{eq:J2}
\begin{align}
\label{eq:J2a}
  & J^a_\S = M^a - \nabla_{\!b}S^{ba}\,,\\[5pt]
  & K^\mu_\S = n_\lambda n_\alpha S^{\lambda\alpha\mu}\,,
\end{align}
\end{subequations}
where $M^a e^\mu_a = n_\lambda M^{\lambda\mu}$ as before, and $e^\mu_a e^\nu_b S^{ab} = n_\lambda S^{\lambda\alpha\beta} P^\mu_\alpha P^\nu_\beta$. Equation~(\ref{eq:J2a}) shows that for strong electromagnetic fields beyond the leading order in the derivative expansion, the surface current is not determined solely by the bulk bound current any more, even in the absence of charged degrees of freedom on the boundary. 
Similarly, the energy-momentum tensor will differ from the simple form (\ref{eq:Tmn-EM00}) beyond leading order.

\subsection{Weak electromagnetic fields}
\label{sec:41}
For weak electromagnetic fields with $A_\mu\sim O(1)$, $g_{\mu\nu}\sim O(1)$, we have the following non-vanishing scalars at order $O(\partial)$ in the derivative expansion:
\begin{subequations}
\label{eq:scalars-1-weak}
\begin{align}
   d=1: \ \ & s^{\scriptscriptstyle (1)} = \left\{B,\ \Omega \right\}\,,\\[5pt]
   d=2: \ \ & s^{\scriptscriptstyle (1)} = \left\{B,\ \Omega \right\}\,,\\[5pt]
   d=3: \ \ & s^{\scriptscriptstyle (1)} = \left\{ \right\}\,.
\end{align} 
\end{subequations}
In $d=1$, we define the vorticity as $\Omega \equiv \epsilon^{\mu\nu}\nabla_{\!\mu} u_\nu = \epsilon^{\mu\nu}a_\mu u_\nu$.
In $d=3$, both the magnetic field and the vorticity are vectors, and there are no scalars at order $O(\partial)$. 

Focussing on $d=2$, the equilibrium generating functional is given by Eq.~(\ref{eq:W0}), with the free energy density 
$$
  {\cal F}(T,\mu,B,\Omega)=P + \MB B + \MO\Omega\,
$$
where $P$, $\MB$,  $\MO$ are functions of $T$ and~$\mu$. Note that both $B$ and $\Omega$ are pseudo-scalars, hence we are describing thermodynamics of a microscopic system which intrinsically violates parity. The bulk current is given by (\ref{eq:JJ}), with the polarization tensor $M^{\alpha\beta} = -\MB\, u_\mu \epsilon^{\mu\alpha\beta}$. The boundary current is $J^a_\S = M^a$, where $M^a e^\mu_a = -\MB\, \epsilon^{\mu\alpha\beta} u_\alpha n_\beta$. 

The bulk energy-momentum tensor can be expressed as a general decomposition (\ref{eq:Tmn0}), whose coefficients are
\begin{align*}
  & {\cal P} = P\,,\\[5pt]
  & {\cal E} = -{\cal F}  + T\frac{\partial {\cal F}}{\partial T}
    + \mu \frac{\partial {\cal F}}{\partial\mu} - \MO\Omega\,,\\[5pt]
  & {\cal Q}^\mu = 
    g_{1}\, \epsilon^{\mu\alpha\beta}u_\alpha a_\beta
   +g_{2} \epsilon^{\mu\alpha\beta}u_\alpha E_\beta 
    \,,\\[10pt]
  & {\cal T}^{\mu\nu} = 0\,,
\end{align*}
where $g_1 \equiv (2\MO - T\frac{\partial\MO}{\partial T} - \mu \frac{\partial\MO}{\partial\mu})$, $g_2\equiv (-\MB+\frac{\partial\MO}{\partial\mu}$), see Ref.~\cite{Jensen:2012jh, Jensen:2011xb}.
The boundary energy-momentum tensor can be expressed in terms of the vector $\ell^\mu\equiv \MO\, \epsilon^{\mu\alpha\beta}n_\alpha u_\beta$ which is tangent to the boundary,
\begin{align}
   T^{ab}_\S = L\gamma^{ab} + \ell^a u^b + \ell^b u^a \,,
\end{align}
where again $\ell^a e^\mu_a = \MO\, \epsilon^{\mu\alpha\beta}n_\alpha u_\beta$, and we have assumed that the surface tension $L$ is constant. The other boundary momentum currents are $\Pi^a_\S = u_n \ell^a$, and $\Pi_\S = 0$. The vector $\ell^a$  (the energy-momentum analogue of the boundary magnetization current) is the density of momentum flowing along the boundary in equilibrium, as is generically expected to happen in a parity-violating system. To sum up, the boundary current is determined by the magnetization $\MB$, while the boundary momentum is determined by $\MO$.

\subsection{Strong electromagnetic fields: 1+1 dimensions}
Now let us turn to strong electromagnetic fields, with $F_{\mu\nu}\sim O(1)$, $g_{\mu\nu}\sim O(1)$. In 1+1 dimensions, there are only two independent scalars in equilibrium at $O(\partial)$ in the derivative expansion, which may be taken to be
$$
  s^{\scriptscriptstyle (1)}_n = 
  \{ \epsilon^{\mu\nu}\nabla_{\!\mu} u_\nu ,\ \epsilon^{\mu\nu} u_\nu \partial_\mu B\}\,.
$$
The equilibrium generating functional is given by Eq.~(\ref{eq:W0}), with the free energy density 
$$
  {\cal F}=P + M_1 \epsilon^{\mu\nu}\nabla_{\!\mu} u_\nu + M_2\, \epsilon^{\mu\nu} u_\nu \partial_\mu B\,,
$$
where $P$, $M_1$, $M_2$ are functions of $T$, $\mu$, and $B$. 
The polarization tensor is given by (\ref{eq:MMG}), and the magnetization $m$ is a combination of the thermodynamic derivatives of $M_1$, $M_2$, and the scalars $s^{\scriptscriptstyle (1)}_n$. The boundary currents are given by Eq.~(\ref{eq:J2}), where $S^{\lambda\mu\nu} = M_2 \epsilon^{\mu\nu} \epsilon^{\lambda\sigma} u_\sigma$.  The energy-momentum tensor may be derived in a straightforward way from the definition (\ref{eq:TJ}).

\subsection{Strong electromagnetic fields: 2+1 dimensions}
In two spatial dimensions, there is a large number of $O(\partial)$ scalars. However, equilibrium relations such as (\ref{eq:Tmueq}), (\ref{eq:dueq}) reduce the number of independent non-zero invariants to just eight. One choice of the independent invariants is listed in Table~\ref{tab:T2}, where $v^\mu\equiv \epsilon^{\mu\alpha\beta}E_\alpha u_\beta$. 

\begin{table}
\begin{center}
\def\arraystretch{1.2}
\setlength\tabcolsep{4pt}
\begin{tabular}{|c|c|c|c|c|c|c|c|c|}
 \hline
 \hline
 $n$ & 1 & 2 & 3 & 4 & 5 & 6 & 7 & 8 \\ 
 \hline
 \hline
 $s^{\scriptscriptstyle (1)}_n$
 & $E^\mu \partial_\mu (\frac{B}{T^2})$   
 & $E^\mu \partial_\mu (\frac{E^2}{T^4})$ 
 & $v^\mu \partial_\mu (\frac{B}{T^2})$   
 & $v^\mu \partial_\mu (\frac{E^2}{T^4})$ 
 & $\nabla_{\!\mu} E^\mu$   
 & $E^\mu \partial_{\mu}T$  
 & $v^\mu \partial_{\mu}T$  
 & $\epsilon^{\lambda\mu\nu}u_\lambda \partial_\nu u_\mu$  
 \\
 \hline
  C & $+$ & $-$ & $+$ & $-$ & $-$ & $-$ & $-$ & $+$ \\
  \hline
  P & $-$ & $+$ & $+$ & $-$ & $+$ & $+$ & $-$ & $-$ \\
  \hline 
  T & $-$ & $+$ & $-$ & $+$ & $+$ & $+$ & $+$ & $-$ \\
  \hline
  W & 3 & 3 & 3 & 3 & 3 & n/a & n/a & 1 \\
  \hline
\end{tabular}
\end{center}
\caption{Independent $O(\partial)$ invariants in 2+1 dimensions. The first row in the table is the number of the invariant, and the second row says what the invariant is. The rows labeled C, P, T indicate the eigenvalue of the invariant under charge conjugation, parity, and time reversal, respectively. Parity in 2+1 dimensions is defined as a reflection of one of the spatial coordinates. The row labeled W shows the weight $w$ of the invariant under a local rescaling of the metric; the invariants which do not transform homogeneously are marked as ``n/a''.}
\label{tab:T2}
\end{table}

The table indicates how the invariants transform under charge conjugation, parity, and time reversal. The table also indicates the weight of the invariants under a Weyl rescaling of the metric, $g_{\mu\nu} \to \tilde g_{\mu\nu} = e^{-2\varphi}g_{\mu\nu}$, where $\varphi$ satisfies $V^\mu\partial_\mu \varphi = 0$. A quantity $\Phi$ transforms homogeneously with weight $w$ under the Weyl rescaling if $\Phi\to \tilde\Phi = e^{w\varphi}\Phi$. For a review of Weyl rescaling in relativistic hydrodynamics, see  Ref.~\cite{Rangamani:2009xk}. Temperature $T$, chemical potential $\mu$,  velocity $u^\mu$, and the electric field $E_\mu$ all have $w=1$. The factors of $T^2$ and $T^4$ in the first four invariants in Table~\ref{tab:T2} are inserted in order to ensure that the invariant has a well-defined weight. For the scalars which transform homogeneously, their weight $w$ coincides with their mass dimension. The invariants $s^{\scriptscriptstyle (1)}_6$ and $s^{\scriptscriptstyle (1)}_7$ do not transform homogeneously and can not appear in a conformally invariant generating functional. 

The first five invariants are in general already non-zero in flat space.
For the static Newtonian gravitational field with potential $\varphi$, we have
$s^{\scriptscriptstyle (1)}_6 \sim E_i\partial_i \varphi$,
$s^{\scriptscriptstyle (1)}_7 \sim \epsilon^{ij} E_i \partial_j \varphi$.
The last invariant is the vorticity, $s^{\scriptscriptstyle (1)}_8 = \Omega$, which is non-zero if the system is rotating. The equilibrium generating functional is then given by Eq.~(\ref{eq:W0}), with the free energy density 
$$
  {\cal F}=P(T,\mu,B,E^2) 
  + \sum_{n=1}^8 M_n(T,\mu,B,E^2)\, s^{\scriptscriptstyle (1)}_n \,.
$$
There are eight scalar functions $M_n$, in addition to pressure, which specify the thermodynamic response at first order. For a system whose microscopic dynamics is PT-invariant, the coefficients $M_3$, $M_4$, and $M_7$ must vanish, in order for the generating functional to be PT-invariant (none of the leading-order invariants are PT-odd). For a system whose microscopic dynamics is conformally invariant, the generating functional must be conformally invariant as well, hence the coefficients $M_6$ and $M_7$ must vanish.

While eight might seem like a large number, if one were to naively write down the constitutive relations directly for $T^{\mu\nu}$ and $M^{\mu\nu}$ in terms of all available $O(\partial)$ scalars, vectors, and tensors, doing so would involve introducing many more than eight unknown $O(1)$ scalar functions, even in equilibrium. The generating functional, on the other hand, allows one to obtain the simplest expressions for the equilibrium quantities without over-counting the parameters.

At leading order in the derivative expansion, the electric polarization vector $p^\lambda$ was simply proportional to the external electric field. At first order, electric polarization can also be induced by the gradients of $T$, $B$, and $E^2$. One finds
\begin{align}
\label{eq:p2}
  p^\lambda = \aE E^\lambda 
  + \gamma_1 \nabla^\lambda T + \gamma_2 \nabla^\lambda B + \gamma_3 \nabla^\lambda E^2
  +\epsilon^{\lambda\rho\sigma}u_\rho 
  \left(
    \gamma_4\, \partial_\sigma T 
  + \gamma_5\, \partial_\sigma B 
  + \gamma_6\, \partial_\sigma E^2
  \right)\,.
\end{align}
The susceptibility here is a function of the parameters $M_n$ of the generating functional,
$$
  \aE = 2\frac{\partial P}{\partial E^2}
  +2\sum_{n=1}^8 \frac{\partial M_n}{\partial E^2} s^{\scriptscriptstyle (1)}_n
  -\frac{\partial M_5}{\partial\mu}
  -\frac{2}{T^4}\nabla_{\!\lambda}
  \left(M_2 E^\lambda + M_4 \epsilon^{\lambda\alpha\beta}E_\alpha u_\beta \right)\,,
$$
and the other coefficients are as follows:
\begin{align*}
 & \gamma_1 = M_6 - \frac{\partial M_5}{\partial T} 
         - \frac{\mu}{T}\frac{\partial M_5}{\partial\mu}
         - \frac{2M_1 B}{T^3} - \frac{4M_2 E^2}{T^5}\,,\\[5pt]
 & \gamma_2 = \frac{M_1}{T^2} - \frac{\partial M_5}{\partial B}\,,\ \ \ \ 
   \gamma_3 = \frac{M_2}{T^4} - \frac{\partial M_5}{\partial E^2}\,,\\[5pt]
 & \gamma_4 = M_7 - \frac{2M_3 B}{T^3} - \frac{4 M_4 E^2}{T^5}\,,\ \ \ \
   \gamma_5 = \frac{M_3}{T^2}\,,\ \ \ \ 
   \gamma_6 = \frac{M_4}{T^4}\,.
\end{align*}
The magnetization is
$$
  m = \frac{\partial P}{\partial B}
  + \sum_{n=1}^8 \frac{\partial M_n}{\partial B} s^{\scriptscriptstyle (1)}_n
  - \frac{1}{T^2} \nabla_{\!\lambda}
  \left(M_1 E^\lambda + M_3 \epsilon^{\lambda\alpha\beta}E_\alpha u_\beta \right)\,.
$$
The above $m$ and $p^\mu$ give the polarization tensor according to Eq.~(\ref{eq:MMG-2}), and thus determine the $O(\partial^2)$ contributions to equilibrium bound charges and bound currents, following (\ref{eq:JJ}).
Finally, we note that the polarization ambiguities of Sec.~\ref{sec:ambiguities} allow one to simplify  the polarization vector $p^\lambda$: adding to the free energy the $W_{\!\varnothing}$ term with $\partial C/\partial T = -\gamma_4$, $\partial C/\partial B = -\gamma_5$, $\partial C/\partial E^2=-\gamma_6$ eliminates the $\gamma_4$, $\gamma_5$, $\gamma_6$ contributions in (\ref{eq:p2}), and adds the term $\partial C/\partial\mu\, \epsilon^{\lambda\alpha\beta}u_\alpha E_\beta$.

The energy-momentum tensor can be read off from the definition (\ref{eq:TJ}), however the general expressions are rather cumbersome, involving thermodynamic derivatives of all eight $M_n$'s. It is easy to derive the energy-momentum tensor when the external electric field vanishes (in a certain set of coordinates), in which case $T^{\mu\nu}$ is only determined by $P$ and $M_8$ (the electric polarization vector is still determined by $M_1$, $M_3$, $M_5$, $M_6$, and $M_7$). In the hydrodynamic-like decomposition (\ref{eq:Tmn0}), the components of the energy-momentum tensor are
\begin{subequations}
\label{eq:T22}
\begin{align}
  & {\cal E} = -{\cal F} 
    + T\frac{\partial{\cal F}}{\partial T}
    + \mu\frac{\partial{\cal F}}{\partial \mu}
    -\MO \Omega \,,\\[5pt]
  & {\cal P} = {\cal F}
    - B\frac{\partial{\cal F}}{\partial B}
    -\MO \Omega\,,\\[5pt]
\label{eq:T22c}
  & {\cal Q}^\mu = g_1\, \epsilon^{\mu\rho\sigma} u_\rho a_\sigma
    + \frac{\partial\MO}{\partial B} 
    \epsilon^{\mu\rho\sigma} u_\rho \partial_\sigma B\,,\\[5pt]
  & {\cal T}^{\mu\nu} = 0\,,
\end{align}
\end{subequations}
where we have defined $\MO\equiv M_8$ and $g_1 \equiv (2\MO - T\frac{\partial\MO}{\partial T} - \mu \frac{\partial\MO}{\partial\mu})$, to mimic the notation in Sec.~\ref{sec:41}, and ${\cal F} = P(T,\mu,B) + \MO(T,\mu,B)\Omega$ after we have set the electric field to zero. Even in flat space and without external electric fields, there is an equilibrium energy flux, caused by the inhomogeneous magnetic field. The magnetization $m$, which determines the spatial bound current according to Eq.~(\ref{eq:JJ23}), simplifies to $m=\partial P/\partial B + \Omega\, \partial\MO/\partial B$.
There is a surface momentum $\ell^a$ flowing along the boundary in equilibrium, completely analogous to the expression in Sec.~\ref{sec:41},
\begin{align}
\label{eq:Ts22}
   T^{ab}_\S = L\gamma^{ab} + \ell^a u^b + \ell^b u^a \,,
\end{align}
where $\ell^a e^\mu_a = \MO(T,\mu,B)\, \epsilon^{\mu\alpha\beta}n_\alpha u_\beta$ may now depend on the external magnetic field.
The other boundary momentum currents are $\Pi^a_\S = u_n \ell^a$, and $\Pi_\S = 0$, as before. The energy-momentum tensors (\ref{eq:T22}), (\ref{eq:Ts22}) will receive extra contributions proportional to the external electric field when the latter is non-zero.

\subsection{Strong electromagnetic fields: 3+1 dimensions}
In 3+1 dimensions, there is again a large number of $O(\partial)$ scalars, but many are not independent due to  equilibrium constraints such as (\ref{eq:Tmueq}), (\ref{eq:dueq}). I counted twenty-one independent non-zero invariants. One choice is listed in Table~\ref{tab:T3}, where $S^\mu = \epsilon^{\mu\rho\sigma\lambda}u_\rho E_\sigma B_\lambda$ is the Poynting vector, $a_\mu = -\partial_\mu T/T$ is the acceleration, and $\Omega^\mu = \epsilon^{\mu\nu\alpha\beta} u_\nu \nabla_{\!\alpha} u_\beta$ is the vorticity. The notation in the table is the same as in the 2+1 dimensional case. The linear combinations in $s^{\scriptscriptstyle (1)}_8$, $s^{\scriptscriptstyle (1)}_9$, and $s^{\scriptscriptstyle (1)}_{10}$ are taken so that the invariant has a well-defined weight under Weyl rescaling. 

\begin{table}
\begin{center}
\def\arraystretch{1.2}
\setlength\tabcolsep{4pt}
\begin{tabular}{|c|c|c|c|c|c|c|c|}
 \hline
 \hline
 $n$ & 1 & 2 & 3 & 4 & 5 & 6 & 7  \\ 
 \hline
 \hline
 $s^{\scriptscriptstyle (1)}_n$
 & $E^\mu \partial_\mu (\frac{B^2}{T^4})$   %
 & $E^\mu \partial_\mu (\frac{E^2}{T^4})$   %
 & $E^\mu \partial_\mu (\frac{E{\cdot}B}{T^4})$  %
 & $B^\mu \partial_\mu (\frac{B^2}{T^4})$   %
 & $B^\mu \partial_\mu (\frac{E^2}{T^4})$   %
 & $B^\mu \partial_\mu (\frac{E{\cdot}B}{T^4})$  %
 & $ \nabla_{\!\mu} (T E^\mu)$
 \\
 \hline
  C & $-$ & $-$ & $-$ & $-$ & $-$ & $-$ & $-$  \\
 \hline
  P & $+$ & $+$ & $-$ & $-$ & $-$ & $+$ & $+$  \\
  \hline 
  T & $+$ & $+$ & $-$ & $-$ & $-$ & $+$ & $+$  \\
  \hline
  W & 3 & 3 & 3 & 3 & 3 & 3 & 4  \\
  \hline
\end{tabular}
\end{center}

\begin{center}
\def\arraystretch{1.2}
\setlength\tabcolsep{4pt}
\begin{tabular}{|c|c|c|c|}
 \hline
 \hline
 $n$ &  8 & 9 & 10   \\ 
 \hline
 \hline
 $s^{\scriptscriptstyle (1)}_n$
 & $B^\mu B^\nu \nabla_{\!\mu} E_\nu {-}B^2 E{\cdot}a {+}3(E{\cdot}B) B{\cdot}a$ %
 & $S^\mu B^\nu \nabla_{\!\mu} E_\nu {+} 2(E{\cdot}B) S{\cdot}a$   %
 & $E^\mu  S^\nu \nabla_{\!\mu}B_\nu {+} (E{\cdot}B) S{\cdot}a$  %
 \\
 \hline
  C &  $-$ & $+$ & $+$  \\
  \hline
  P &  $+$ & $-$ & $-$   \\
  \hline 
  T &  $+$ & $+$ & $+$  \\
  \hline
  W &  7 & 9 & 9 \\
  \hline
\end{tabular}
\end{center}

\begin{center}
\def\arraystretch{1.2}
\setlength\tabcolsep{4pt}
\begin{tabular}{|c|c|c|c|c|c|}
 \hline
 \hline
 $n$ & 11 & 12 & 13 & 14 & 15 \\ 
 \hline
 \hline
 $s^{\scriptscriptstyle (1)}_n$
 & $S^\mu \partial_\mu (\frac{B^2}{T^4})$   %
 & $S^\mu \partial_\mu (\frac{E^2}{T^4})$   %
 & $S^\mu \partial_\mu (\frac{E{\cdot}B}{T^4})$  %
 & $ \nabla_{\!\mu} (\frac{S^\mu}{T})$
 & $\epsilon^{\mu\nu\rho\sigma} u_\mu B_\nu \nabla_{\!\rho} B_\sigma$ 
 \\
 \hline
  C & $+$ & $+$ & $+$ &  $+$ & $+$ \\
  \hline
  P & $+$ & $+$ & $-$ &  $+$ & $-$ \\
  \hline 
  T & $-$ & $-$ & $+$ &  $-$ & $+$ \\
  \hline
  W & 5 & 5 & 5 & 4 & 5 \\
  \hline
\end{tabular}
\end{center}
\begin{center}
\def\arraystretch{1.2}
\setlength\tabcolsep{4pt}
\begin{tabular}{|c|c|c|c|c|c|c|}
 \hline
 \hline
 $n$  & 16 & 17 & 18 & 19 & 20 & 21  \\ 
 \hline
 \hline
 $s^{\scriptscriptstyle (1)}_n$
 & $E{\cdot}a$   %
 & $B{\cdot}a$  %
 & $S{\cdot}a$   %
 & $E{\cdot}\Omega$   %
 & $B{\cdot}\Omega$  %
 & $S{\cdot}\Omega$
 \\
 \hline
  C  & $-$ & $-$ & $+$ & $-$ & $-$ & $+$  \\
  \hline
  P  & $+$ & $-$ & $+$ & $-$ & $+$ & $-$  \\
  \hline 
  T  & $+$ & $-$ & $-$ & $-$ & $+$ & $+$  \\
  \hline
  W  & n/a & n/a & n/a & 3 & 3 & 5  \\
  \hline
\end{tabular}
\end{center}

\caption{Independent $O(\partial)$ invariants in 3+1 dimensions.
}
\label{tab:T3}
\end{table}

The first fifteen invariants are in general non-zero already in flat space. The equilibrium generating functional is given by Eq.~(\ref{eq:W0}), with the free energy density 
\begin{equation}
\label{eq:F2}
  {\cal F}=P(T,\mu,E^2, B^2, E{\cdot}B) 
  + \sum_{n=1}^{21} M_n(T,\mu,E^2, B^2, E{\cdot}B)\, s^{\scriptscriptstyle (1)}_n \,.
\end{equation}
There are twenty-one scalar functions $M_n$, in addition to pressure, which specify the thermodynamic response at first order. For a system whose microscopic dynamics is PT-invariant, the coefficients $M_9, \dots, M_{15}$, $M_{18}$, and $M_{21}$ must vanish, in order for the generating functional to be PT-invariant (none of the leading-order invariants are PT-odd). For a system whose microscopic dynamics is conformally invariant, the generating functional must be conformally invariant as well, hence the coefficients $M_{16}$, $M_{17}$, and $M_{18}$ must vanish. 

Analogously to what happens in 2+1 dimensions, polarization may be induced by the gradients of the applied fields. The electric polarization vector which follows from the free energy is 
\begin{align}
\label{eq:p3}
  p^\lambda & = \aEE E^\lambda + \aEB B^\lambda 
  + \chi_{\rm\scriptscriptstyle E\Omega} \Omega^\lambda 
  + \chi_{\rm\scriptscriptstyle ES} S^\lambda \nonumber\\[7pt]
  & + \gamma_1 \nabla^\lambda T + \gamma_2 \nabla^\lambda B^2
    + \gamma_3 \nabla^\lambda E^2 + \gamma_4 \nabla^\lambda (E{\cdot}B) \nonumber\\[7pt]
  & + \gamma_5 X^{\lambda\alpha}\partial_\alpha T 
    + \gamma_6 X^{\lambda\alpha}\partial_\alpha B^2
    + \gamma_7 X^{\lambda\alpha}\partial_\alpha E^2
    + \gamma_8 X^{\lambda\alpha}\partial_\alpha (E{\cdot}B)
    + \gamma_9 X^{\lambda\alpha} \Omega_\alpha \nonumber\\[7pt]
  & + \gamma_{10} \Delta^\lambda_\rho S^\alpha \nabla_{\!\alpha} B^\rho
    + \gamma_{11} \Delta^\lambda_\rho S^\alpha \nabla^\rho B_\alpha
    + \gamma_{12} \Delta^\lambda_\rho B^\alpha \nabla_{\!\alpha} B^\rho \nonumber\\[7pt]
  & + \gamma_{13} X^{\lambda\rho} B^\sigma \nabla_{\!\rho} E_\sigma
    + \gamma_{14} X^{\lambda\rho} E^\sigma \nabla_{\!\sigma} B_\rho \,,
\end{align}
where $X^{\alpha\beta}\equiv \epsilon^{\alpha\beta\rho\sigma}u_\rho B_\sigma$.
The susceptibility coefficients $\aEE$ etc and $\gamma_k$ are determined by thermodynamic derivatives of the coefficients $M_n$, and can be easily read off from the free energy density (\ref{eq:F2}).
However, as explained in Sec.~\ref{sec:ambiguities}, polarization vectors only make sense up to certain redefinitions. For example, by choosing the arbitrary vector $C_\mu$ in (\ref{eq:pm-shift}) appropriately, one can eliminate $\chi_{\rm\scriptscriptstyle E\Omega}$, and trade $\chi_{\rm\scriptscriptstyle ES}$, $\gamma_5$, $\gamma_6$, $\gamma_7$, $\gamma_8$ in favor of a single contribution proportional to $\epsilon^{\lambda\nu\rho\sigma}u_\nu \nabla_{\!\rho} B_\sigma$. The coefficients $\aEE$ and $\aEB$ suffer from similar ambiguities.

The magnetic polarization vector which follows from the free energy is 
\begin{align}
  m^\lambda & = \aBB B^\lambda + \aBE E^\lambda 
  + \chi_{\rm\scriptscriptstyle B\Omega} \Omega^\lambda 
  + \chi_{\rm\scriptscriptstyle BS} S^\lambda \nonumber\\[7pt]
  & + \delta_1 \nabla^\lambda T + \delta_2 \nabla^\lambda B^2
    + \delta_3 \nabla^\lambda E^2 + \delta_4 \nabla^\lambda (E{\cdot}B) \nonumber\\[7pt]
  & + \delta_5 Y^{\lambda\alpha}\partial_\alpha T 
    + \delta_6 Y^{\lambda\alpha}\partial_\alpha B^2
    + \delta_7 Y^{\lambda\alpha}\partial_\alpha E^2
    + \delta_8 Y^{\lambda\alpha}\partial_\alpha (E{\cdot}B)
    + \delta_9 Y^{\lambda\alpha} \Omega_\alpha \nonumber\\[7pt]
  & + \delta_{10} \Delta^\lambda_\rho B^\mu \nabla^\rho E_\mu
    + \delta_{11} \Delta^\lambda_\rho B^\mu \nabla_{\!\mu} E^\rho
    + \delta_{12} Y^{\lambda\rho} B^\sigma \nabla_{\!\rho} E_\sigma
    + \delta_{13} Y^{\lambda\rho} E^\sigma \nabla_{\!\sigma} B_\rho \nonumber\\[7pt]
  & + \delta_{14} \Delta^\lambda_\rho S^\mu \nabla_{\!\mu} E^\rho
    + \delta_{15} \Delta^\lambda_\rho E^\mu \nabla_{\!\mu} S^\rho
    + \delta_{16} \epsilon^{\lambda\mu\rho\sigma} u_\mu \nabla_{\!\rho} B_\sigma\nonumber\\[7pt]
  & + \delta_{17} X^{\lambda\alpha}\partial_\alpha T 
    + \delta_{18} X^{\lambda\alpha}\partial_\alpha B^2
    + \delta_{19} X^{\lambda\alpha}\partial_\alpha E^2
    + \delta_{20} X^{\lambda\alpha}\partial_\alpha (E{\cdot}B)\,,
\end{align}
where $Y^{\alpha\beta}\equiv \epsilon^{\alpha\beta\rho\sigma}u_\rho E_\sigma$.
The susceptibility coefficients $\aBB$ etc and $\delta_k$ can be easily read off from the free energy density (\ref{eq:F2}). The ambiguities (\ref{eq:pm-shift}) also affect the magnetic polarization: adding to the free energy the $W_{\!\varnothing}$ term with $C^\mu = C u^\mu$ shifts $\aBE \to \aBE + \partial C/\partial\mu$ (in addition to shifting $\delta_1$, $\delta_2$, $\delta_3$, $\delta_4$). 

While the polarization vectors are ambiguous, the energy-momentum tensor and the current are not. As an example, consider the $M_{21}$ term in the free energy. It gives rise to polarization vectors $p^\mu = M_{21} \epsilon^{\mu\nu\rho\sigma}u_\nu B_\rho \Omega_\sigma$ and $m^\mu = -M_{21} \epsilon^{\mu\nu\rho\sigma} u_\nu E_\rho \Omega_\sigma$ which do not suffer from polarization ambiguities. Such contributions to $p^\mu$ and $m^\mu$ only come from $M_{21}$, and therefore the magneto-vortical response of the surface charge density is correlated with the electro-vortical response of the surface current. The corresponding boundary current is 
$$
  J^\mu_\S = \left(M_{21} n_\alpha \epsilon^{\alpha\beta\rho\sigma}u_\beta B_\rho \Omega_\sigma \right) u^\mu + M_{21}\left( \Omega_n E^\mu - E_n \Omega^\mu \right)\,,
$$
where $\Omega_n\equiv \Omega{\cdot}n$, $E_n\equiv E{\cdot}n$, and we have assumed $u_n=0$. 

As another example, let us set the electric field to zero (in a certain set of coordinates), while keeping the magnetic field non-zero. The equilibrium bulk energy-momentum tensor is then determined by only four functions $M_4$, $M_{15}$, $M_{17}$, and $M_{20}$, in addition to the leading-order pressure $P=P(T,\mu, B^2)$. The correction to the leading-order energy-momentum tensor (\ref{eq:Tmn-3}) is straightforward to derive, and we will not write it down explicitly. Both ${\cal E}$, ${\cal P}$, and ${\cal T}^{\mu\nu}$ will receive derivative corrections, proportional to $\partial_\mu T$, $\nabla_{\!\mu} B_\nu$, and $\Omega^\mu$. In addition, the magneto-vortical term $M_{20}$ will give rise to equilibrium energy currents ${\cal Q}^\mu$ proportional to $ \epsilon^{\mu\nu\rho\sigma}u_\nu B_\rho a_\sigma$, and $ \epsilon^{\mu\nu\rho\sigma}u_\nu B_\rho \partial_\sigma B^2$ .
There is also a non-vanishing boundary energy-momentum tensor $T^{\alpha\beta}_\S$, defined by (\ref{eq:TJ}). In the decomposition $T^{\alpha\beta}_\S = {\cal E}_\S u^\alpha u^\beta + {\cal P}_{\!\S}\, \Delta^{\alpha\beta} +({\cal Q}_\S^\alpha u^\beta + {\cal Q}_\S^\beta u^\alpha) + {\cal T}_\S^{\alpha\beta}$ as in (\ref{eq:Tmn0}) we have
\begin{subequations}
\begin{align}
  & {\cal E}_\S = -B{\cdot}n \left( M_{17} + 4M_4 B^2/T^4 \right)\,,\\[5pt]
  & {\cal P}_\S = -\coeff43 B{\cdot}n\, M_4 B^2/T^4\,,\\[5pt]
  & {\cal Q}_\S^\alpha = M_{20}\, n_\rho B_\mu u_\nu\, \epsilon^{\rho\mu\nu\alpha}\,,\\[5pt]
  & {\cal T}_\S^{\alpha\beta} = 2 B{\cdot}n\, M_4/T^4 
    \left( B^\alpha B^\beta - \coeff13 \Delta^{\alpha\beta} B^2 \right)\nonumber\\
  &\ \ \ \ \ \  + M_{15} n_\rho u_\mu B_\nu 
    \left( \epsilon^{\rho\mu\nu\alpha}B^\beta + \epsilon^{\rho\mu\nu\beta}B^\alpha\right)\,.
\end{align}
\end{subequations}
Here $n^\mu$ is the unit normal vector the boundary as before, and we have omitted the surface tension $L$. One can see  that in addition to the standard surface tension, even a uniform magnetic field generates energy density, pressure, energy current, and spatial stress on the boundary.

\section{Summary}
Let us summarize. We have presented the equilibrium free energy of isotropic relativistic matter, in the regime when external electromagnetic fields are more important than external gravitational fields. From a technical point of view, this amounts to generalizing the analysis of Ref.~\cite{Jensen:2012jh} by {\it i)} performing a partial summation of electromagnetic contributions, and {\it ii)} by taking into account surface terms in the generating functional. From a physical point of view, this amounts to describing the effects of polarization. The equilibrium electric current can be expressed in terms of the polarization tensor $M^{\mu\nu}$ to all orders, 
$
  J^\alpha = \rho u^\alpha  
  - \nabla_{\!\lambda} M^{\lambda\alpha}\,.
$
The charge density and the spatial current are given by Eqs.~(\ref{eq:NN23}) and (\ref{eq:JJ23}). In 3+1 dimensions in flat space%
\footnote{
In our construction, the temperature gradient was induced by the external metric $\delta g_{\mu\nu}$, say $T=T_0 + \delta T_g$, where $T_0$ is constant and the arbitrary static $\delta T_g$ is induced by $\delta g_{\mu\nu}$. This gives a contribution to the equilibrium charge density $\delta n_g = - \p {\cdot}\grad \delta T_g /T_0$. If in addition we couple the system to an external heat bath with temperature $\delta T_B = -\delta T_g$, the overall temperature becomes constant, hence there must be a contribution $- \p {\cdot}\grad \delta T_B /T_0$, cancelling $- \p {\cdot}\grad \delta T_g /T_0$. This shows that even in flat space with $\delta T_g = 0$ there is a contribution to the charge density $\delta n_B = - \p {\cdot}\grad T /T$, provided equilibrium is maintained.
}
 they reduce to
\begin{align}
  n & =\rho - \grad{\cdot}\p -  \p {\cdot}\grad T /T
     - 2{\bf m}{\cdot}{\bm \omega}\,,\\[5pt]
  {\bf j} & = \grad{\times}{\bf m} + {\bf m}{\times} \grad T/T\,,
\end{align}
where $\p$ is the electric polarization vector, ${\bf m}$ is the magnetic polarization vector, and ${\bm \omega}$ is the angular velocity. These expressions generalize the familiar $n=\rho - \grad{\cdot}\p$ and ${\bf j} = \grad{\times}{\bf m}$ in electro- and magneto-statics. 
At leading order in the derivative expansion, the surface current is $J^\mu_\S = n_\lambda M^{\lambda\mu}$, which says that the surface charge density is $\p{\cdot}\n$, and the surface spatial current is ${\bf m}{\times}\n$.

The notion of polarization is ambiguous when the external fields vary in space. This is because polarization is defined as a response to electric and magnetic fields, which are not fundamental quantities, but are rather derived from the vector potential~$A_\mu$. In particular, the electro-vortical susceptibility $\chi_{\rm\scriptscriptstyle E\Omega}$ is unphysical, as well as the magneto-electric susceptibility $\aBE$ in the presence of free charges. Nevertheless, most $O(\partial)$ contributions to polarization are not affected by this ambiguity and may be derived from the equilibrium free energy, as described in Sec.~\ref{sec:next-order}. 
For example, in a parity-violating system, there is a contribution to the electric polarization vector $\p \propto {\bf B}\times{\bm\omega}$, and the contribution to the magnetic polarization vector ${\bf m} \propto {\bf E}\times{\bm\omega}$ which do not suffer from this ambiguity. The corresponding surface charge density $\sigma_\S = c\,\n{\cdot}({\bf B}\times{\bm\omega})$ and the surface current ${\bf j}_\S = c\,( {\bf E} ({\bm\omega}{\cdot}\n) - {\bm\omega} ({\bf E}{\cdot}\n))$ are determined by the same coefficient $c=2M_{21}$.

When the external fields are non-uniform, the boundary charge and spatial current are no longer determined by polarization. This is not surprising: while the polarization vectors are ambiguous, the charge and the current are not. 
At first order in the derivative expansion the boundary current is
\begin{equation}
\label{eq:JSS}
   J^\mu_\S = n_\lambda M^{\lambda\mu} - \nabla_{\!\lambda} \! \left( n_\alpha P^\lambda_\rho P^\mu_\sigma S^{\alpha\rho\sigma} \right)\,,
\end{equation}
where $P^\lambda_\rho = \delta^\lambda_\rho - n^\lambda n_\rho$, and $S^{\alpha\rho\sigma}$ is defined by (\ref{eq:Slmn}). For the generating functional (\ref{eq:F2}) in 3+1 dimensions, there are 15 contributions to $S^{\alpha\rho\sigma}$. 
As an example, consider the effect of the $M_2$ term for non-rotating matter at constant temperature in flat space.
For the boundary with vanishing extrinsic curvature ($\nabla_{\!\mu} n_\nu = 0$), the surface charge density is 
$$
  \sigma_\S = \p{\cdot}\n - \frac{2}{T^4} \partial_a(M_2 E^a E_{n})\,,
$$
where $E_{n}$ is the normal component of the electric field. This describes the response of the boundary charge density to the changes of the external electric field along the boundary.

The equilibrium energy-momentum tensor to leading order takes a simple form (\ref{eq:Tmn-EM00}) which we repeat here:
\begin{subequations}
\label{eq:Tmn-EM-disc}
\begin{equation}
  T^{\mu\nu} = P g^{\mu\nu} + (Ts + \mu  \rho) u^\mu u^\nu
  + T^{\mu\nu}_\EM\,,
\end{equation}
with
\begin{equation}
  T^{\mu\nu}_\EM = M^{\mu\alpha} g_{\alpha\beta} F^{\beta\nu} 
  + u^\mu u^\alpha \left(
  M_{\alpha\beta} F^{\beta\nu} - F_{\alpha\beta} M^{\beta\nu} \right)\,.
\end{equation}
\end{subequations}
This expression is model-independent, and is a leading-order consequence of  gauge invariance,  diffeomorphism invariance, and locality (on scales longer than the screening length). Beyond the leading order in derivatives, the form of the equilibrium $T^{\mu\nu}$ is more complicated.

Equilibrium $\mu$-independent contributions to the current $J^\alpha$ are usually referred to as ``bound charges'' and ``bound currents''. There exist analogous contributions to the equilibrium $T^{\alpha\beta}$, which one may similarly christen ``bound energy'', ``bound pressure'', ``bound momentum'', and ``bound stress''. Just like bound charges and bound currents, these live both in the bulk and on the surface. For matter subject to external magnetic field (and no electric field), there will be bulk energy currents ${\bf Q}\propto {\bf B} \times \grad T$, ${\bf Q}\propto {\bf B} \times \grad B^2$. The boundary energy current ${\bf Q}_\S = \chi_{\rm\scriptscriptstyle B\Omega}\, {\bf B}\times \n$ is determined by the same susceptibility $\chi_{\rm\scriptscriptstyle B\Omega}$ which fixes the response of magnetization to rotation, ${\bf m} = \aBB {\bf B} + 2 \chi_{\rm\scriptscriptstyle B\Omega} {\bm\omega} + O(\partial T,\partial B)$.

Finally, our discussion so far was restricted to the state of global equilibrium, i.e.\ to thermodynamics. It is straightforward to extend it to {\em hydrodynamics} of polarized relativistic matter, if one assumes that the external electromagnetic and gravitational fields are not dynamical. In order to do so, one promotes $u^\mu$ and $T$ to dynamical variables, and postulates the hydrodynamic equations in the form $\nabla_{\!\mu} T^{\mu\nu} = F^{\nu\lambda} J_\lambda$, $\nabla_{\!\mu}J^\mu = 0$, with the leading-order energy-momentum tensor given by Eq.~(\ref{eq:Tmn-EM-disc}). Beyond the leading order, the energy-momentum tensor becomes  much more involved as discussed in Sec.~\ref{sec:next-order}, plus the extra transport coefficients such as viscosity make their way into the hydrodynamic equations. 

If the electromagnetic fields are dynamical, the conservation equations for $T^{\mu\nu}$ and $J^\mu$ need to be supplemented by the evolution equations for the electromagnetic fields. These are usually taken to be Maxwell's equations, $\nabla_{\!\nu}F^{\mu\nu} = J^\mu$. Substituting the equilibrium current (\ref{eq:JJ}) gives
\begin{equation}
\label{eq:ME}
  \nabla_{\!\nu}\left( F^{\mu\nu} - M^{\mu\nu} \right) = \rho u^\mu\,,
\end{equation}
which is the standard covariant form of Maxwell's equations in matter, see e.g.~\cite{deGroot}. In the right-hand side of (\ref{eq:ME}), $\rho$ is the density of free charges, while the effects of polarization are in the left-hand side. 
In the framework of (\ref{eq:ME}), the derivative expansion in hydrodynamics can be implemented through the derivative expansion for $M^{\mu\nu}$, however Eq.~(\ref{eq:ME}) itself will receive corrections, e.g.\ due to the electrical conductivity. We plan to return to the study of hydrodynamics of polarized relativistic matter in the future.

\subsection*{Acknowledgments}
I would like to thank Kristan Jensen and Adam Ritz for helpful conversations. This work was supported in part by NSERC of Canada.

\bibliographystyle{JHEP}
\bibliography{tprm}

\end{document}